\documentclass[sigconf]{acmart}
\AtBeginDocument{%
  \providecommand\BibTeX{{%
    \normalfont B\kern-0.5em{\scshape i\kern-0.25em b}\kern-0.8em\TeX}}}

\copyrightyear{2023}
\acmYear{2023}
\setcopyright{rightsretained}
\acmConference[ICGJ 2023]{International Conference on Game Jams, Hackathons
and Game Creation Events}{August 30, 2023}{Virtual Event, Ukraine}
\acmBooktitle{International Conference on Game Jams, Hackathons and Game
Creation Events (ICGJ 2023), August 30, 2023, Virtual Event,
Ukraine}\acmDOI{10.1145/3610602.3610610}
\acmISBN{979-8-4007-0879-4/23/08}

\acmConference[ICGJ '23]{
  International Conference on Game Jams, Hackathons and Game Creation Events}{August 30,
  2023}{Virtual Event, Ukraine}
\acmBooktitle{International Conference on Game Jams, Hackathons and
Game Creation Events Proceedings } 
\acmPrice{00.00}
\acmISBN{978-1-4503-XXXX-X/18/06}

\begin{document}

\title{Online Quantum Game Jam}

\author{Laura Piispanen}
\orcid{0000-0002-5547-9069}
\affiliation{%
  \institution{Aalto University}
  \streetaddress{kuja}
  \city{Espoo}
  \country{Finland}}
\email{laura.piispanen@aalto.fi}

\author{Daria Anttila}
\orcid{0000-0003-1842-6336}
\affiliation{%
  \institution{University of Turku}
  \streetaddress{osoite}
  \city{Turku}
  \country{Finland}}
\email{daria.anttila@utu.fi
}

\author{Natasha Skult}
\orcid{0009-0005-4728-2281}
\affiliation{%
  \institution{University of Turku}
  \streetaddress{osoite}
  \city{Turku}
  \country{Finland}}
\email{natasha.skult@utu.fi}
\renewcommand{\shortauthors}{Piispanen, Anttila, Skult}

\begin{abstract}
This paper presents and discusses the online version of Quantum Game Jams, events where quantum physics related games are created. It consists of a two-part investigation into online Quantum Game Jams. The first part involves examining three events that took place between 2020 and 2021. The second part provides a detailed account of organising the Global Quantum Game Jam from 2021 to 2022, evaluating its outcomes based on participant feedback and experiences. Additionally, it examines the backgrounds of the participants in the global events of 2021 and 2022. Based on the findings, this paper proposes a set of guidelines for organising future online Quantum Game Jams, which can also be applicable to game jams and science game jams in general.
\end{abstract}




\received{09 May 2023}
\received[accepted]{12 July 2023} 

\maketitle
\section{Introduction}
Game Jams are events for interdisciplinary creation, learning, and specifically game developing under a strict interval and an event-specific theme or restriction \cite{kultima2015, arya2013}. At a \textit{Quantum Game Jam} (QGJ) the theme of the jam is connected to quantum physics or quantum computing and the end-creations are expected to have at least one dimension of a quantum game in them \cite{kultima2021qgj,piispanen2023}. In the context of video games, card games, board games and other games \textit{Quantum Games} are defined as “\textit{games that reference the theory of quantum physics, quantum technologies or quantum computing through perceivable means, connect to quantum physics through a scientific purpose or use quantum technologies.}” \cite{piispanen2023}. 

QGJs have their origin at the intent of creating citizen science games and educational games for and about quantum physics and quantum technologies, and have been able to create a fertile ground for several citizen science game prototypes and art pieces \cite{kultima2021qgj, piispanen2023projects}. For game developers, these events offer important opportunities for working in a multidisciplinary project. The first QGJ was organised in 2014 by the Finnish Game Jam organisation together with the quantum technology focused research group Turku Quantum Technology (TQT) from Turku University. To date, more than 270 games that incorporate quantum physics concepts have been developed, with 126 of them originating from game jams and 33 from hackathons \cite{quantumgames}. 

The onset of the Covid-19 pandemic in early 2020 necessitated a shift to remote interactions for events, work, and public services, among other sectors of life. As a response, many game jams turned online, which was also the case for QGJs as the first online QGJ was organised in April 2020 by the student initiative quantum computing community IndiQ. In October 2020 two QGJs were organised. First by IndiQ and one right after as a joint effort between the Pisa Internet Festival and TQT. Other QGJs and hackathons were organised that year around the world, also mostly online. 
In 2021, a series of online QGJs, under the name of \textit{Global Quantum Game Jam}, began as a result of a collaboration between the \textit{IGDA Finland}, \textit{Aalto University}, and \textit{University of Turku}. This event has since been hosted on a Discord server, which has grown into a community of over 450 members, enthusiasts from disciplines of game development and quantum physics. This paper aims to introduce the key takeaways from the series of online QGJs by providing best practices for organising such events. With this contribution we hope to contribute to the field of science game jams and science game hackathons.\\

This paper presents a two-part investigation. The first part examines the experiences of participants in three online game jams with a quantum physics theme, conducted between 2020 and 2021. The objective was to gather valuable insights to inform the development of the Global QGJ. This investigation aimed to uncover the backgrounds of the participants in terms of game development and quantum physics, as well as their overall experiences during the events. The observations and findings from this study played a crucial role in shaping the event design of the Global QGJ. 

The second phase of the study focuses on investigating the experiences of participants in two iterations of the Global QGJ. The objective is to assess the diversity in terms of nationality and age among participants, as well as their prior involvement in game development and quantum physics. This analysis aims to determine the effectiveness of the event design in attracting and tending for a well-rounded and diverse participant pool. Additionally, the study evaluates the use of the learning resources provided. It further examines participants' reported insights gained from the event, and their overall event experiences in order to define the main features of the event design supporting a global, online co-creation in game development. 

The games developed in these iterations of the Global QGJ are analysed separately to identify their connection to quantum physics using the definition for quantum games by Piispanen et al. (2023) \cite{piispanen2023}.
\section{Game Jams and quantum games}
Game jams are events meant for game development in a supportive and open environment for people from varying backgrounds and have been organised since 2002 \cite{kultima2015, kultima2018, kultima2016fgj, laiti2020, lai2021}. Much like in a jam session, where musical acts are produced in a collaborative manner with little or no preparation prior to the event, at a game jam, game prototypes are being designed and developed from start to finish. A game jam usually has constraints related to the duration of the event, the theme, the technology used and sometimes even the location \cite{kultima2015,lai2021}. Game jams are being organised by various actors such as companies, organisations, educational institutes and individuals enthusiastic to use game development tools for creating games or other interactive content. In the game industry, the common practice among game development companies is to organise internal game jams when planning possible new commercial game prototypes \cite{musil2010}.

Many game jams follow the structure of the large-scale game jams with the longest history, the Global Game Jam\footnote{\url{https://globalgamejam.org/}} or the online game jam track of \textit{Ludum Dare}\footnote{\url{https://ldjam.com/}}, which include a 48 hour limit for developing under a given theme. Game jams are able to facilitate collaboration between creative individuals from unexpected disciplines, who might not otherwise step into game creation \cite{kultima2015, kultima2021qgj}. 

On-site game jams have become popular phenomena not only for game development and the facilitated creative process, but also for networking and socialising \cite{zook2013,ho14,kultima2016fgj, kultima2021qgj}, \cite{turner2013, reng2013}. However, at the beginning of 2020, many game jams were compelled to transition to online formats due to a global pandemic. This transition brought significant changes to many aspects, including challenges in fostering socialisation, communication, and keeping up with long-standing traditions, as well as limitations imposed by online platforms \cite{koskinen2021}. While events like \textit{Ludum Dare} had already been organised online for years, many on-site jams had to justify their shift to virtual environments. \\

Teaching and learning opportunities are another crucial aspect of game jams \cite{preston2012,kultima2021expert, arya2013, fowler2013learn}. The serious or applied aspects of game development have been extensively explored in the context of game jams, ranging from teaching about accessibility issues \cite{scott2013} to examining how game jams can foster collaborative design processes between scientists and game developers \cite{kultima2021qgj}. This paper contributes to this perspective by facilitating the development of quantum games within the context of science game jams \cite{kultima2021qgj}. By doing so, it continues the continuum of research on leveraging game jams to bridge the gap between science and game development.

\textbf{Quantum games} can be analysed through "three dimensions of quantum games," which include the dimensions of quantum physics, quantum technologies, and scientific purposes \cite{piispanen2023}. These dimensions provide a framework for evaluating whether a game qualifies as a quantum game and facilitate the creation of games that establish meaningful connections to quantum physics. So far, almost half of all the quantum games have been created through quantum physics themed game jams, and as such serve as a benchmark for those who consider making (commercial) quantum games \cite{skult2022, piispanen2023, piispanen2023history, quantumgames}. 
\section{Quantum Game Jam} 
Quantum Game Jam (QGJ) is an interdisciplinary and non-competitive event for creating games under a quantum physics related theme. The event has been organised annually since 2014, when it started as an on-site event with varying, sometimes even peculiar locations, like quantum physics laboratories, planetariums, and a Ferris wheel \cite{kultima2021qgj}. A distinctive feature of QGJs has been, since the beginning, short keynote speeches, that introduce quantum physics and the possible tools provided for incorporating, for example, numerical quantum simulations to the development process. Such tools have been originally designed for citizen science game prototypes for specific research questions or even offered a way to call a quantum computer in the gameplay \cite{kultima2021qgj, piispanen2023projects, piispanen2023history}. Like many game jams, QGJ has also been characterised by distinct time constraints for game development, an inclusive approach that welcomes all participants, and the public sharing of games through a dedicated online platform \cite{kultima2021qgj}.

In 2019 a special kind of invite-only QGJ, the \textit{Quantum Wheel} was organised in collaboration with Finnish Game Jam Association, the Turku Quantum Technology research group (TQT) and IBM. The event provided the participants special, first-time access to the use of quantum computers. The participants consisted of experienced game jammers, award-winning indie developers and quantum physics researchers, and questionnaires about their experiences were collected throughout the 3-day event \cite{kultima2021qgj}. These observations provided valuable learning for starting a new era of QGJs, as they turned fully online in 2020. \\

The first online QGJ studied in this paper was organised in October 2020 by the Indian student-initiative quantum computing community IndiQ on their Discord server and 7 games were submitted under the event page on Itch.io. A week later another online QGJ was organised together by the Italian PISA Internet Festival and the Finnish TQT, where 8\footnote{In addition to these, a game developed in IndiQ QGJ was also submitted.} games were developed during the event. 

QGJs, quantum computing themed hackathons and other multidisciplinary events have gathered the attention of many actors. However, these events are still reaching their audience mainly from the quantum physics research communities or quantum computing communities instead of a much wider audience. It has been noted that in order to develop games with meaningful connections to quantum physics, expertise is needed both in quantum physics as well as game design \cite{kultima2021qgj,archer2022,piispanen2023projects, piispanen2023}. QGJ has been somewhat successful in gathering a balanced participant list of experienced game developers, game jammers and quantum physicists, but until 2021, the physics expertise has mostly consisted of a single group of quantum physics researchers. In addition, the facilitation between these multidisciplinary teams has faced challenges in terms of communication and the absence of shared practices in creative work \cite{kultima2021qgj,archer2022}.

In 2021 \textit{IGDA Finland}, \textit{Aalto University}, and \textit{University of Turku} started a series of online QGJs, under the name of \textit{Global Quantum Game Jam}, where authors were able to modify the structure of the event according to their learnings and study how well the event is able to cater for quantum game development. The event has since the beginning been hosted on a dedicated Discord server, that today has a population of over 650 quantum game enthusiasts. 
\section{Methods}
Since 2017, the authors have organised five Quantum Game Jams (QGJ), and in addition altogether over 30 game jams, from which 12 have been online or hybrid events. For developing a more accessible event, this paper studied the participants and their experiences of online QGJs in two parts.\\
\subsection{Participants of online QGJs}
The first part to this study was conducted as a mixed-methods study to investigate the backgrounds of participants in online QGJ events of the IndiQ Quantum Game Jam 2020 (IndiQ QGJ 2020), IF Quantum Game Jam 2020 (IFQGJ 2020), and the Games Now! 2021 \#4 game jam (GN!2021), that was quantum physics themed. The aim was to understand what were the backgrounds of the participants with regards to game development and quantum physics, how they found the event and whether they’d have any constructive criticism to give. The study consisted of both observation and questionnaire data. Questions are presented in the Appendix. 

The game jams of IndiQ QGJ 2020 and IFQGJ 2020 were one-weekend-long events organised in October 2020 only a week apart from each other, each running from a Friday to the next Sunday with about 48 hours of game development. On the first days of the events after team formation, participants were asked to complete an online questionnaire about their backgrounds with game development and quantum physics. A second set of feedback was collected on the third day of the event and the games were submitted. 

Participants were recruited through the main platforms used in the events, which for all the events was a Discord server set up by the event organisers. Participants joined the study through online links to questionnaires distributed on public channels of the event servers with the permission of the event organisers. The permission to gather demographic information such as participants' age, nationality, or gender was not granted. Quantitative data about the backgrounds of the participants on game development and quantum physics were collected using single- and multiple-choice questions, while qualitative data about their experiences were collected using open-ended questions and observation notes. Very basic descriptive statistics were used to analyse the quantitative data through Google Sheets, while thematic analysis was used to analyse the qualitative data for any themes recurring in the respondents' answers.

The GN!2021 was one game jam from the set of expert-driven game jams each spreading over the duration of one week \cite{kultima2021expert}. For this event one questionnaire was distributed after the event through the event server on discord. Participants were asked to explain their development experiences and overall experiences related to the event retrospectively through open-ended questions. Information on their backgrounds related to game development experience and possible education on quantum physics were gathered through single- and multiple-choice questions. In addition, self-reflected experiences on learning were asked through open-ended questions. \\

In addition to observations made during the three events and the questionnaires distributed, information on the submitted games in these events was gathered from the events' Itch.io page. Notably, a considerable number of these games did not provide information about the creators. The possible connection to quantum physics was examined from the gameplay and from their description on the Itch.io pages. Additional insight to part of the games were gained through examination of the open-ended questions. The connection was examined through the three dimensions of quantum games \cite{piispanen2023}. 
\subsection{Evaluating the Global Quantum Game Jam}
The analysis conducted in the first segment served as a foundation for making well-informed decisions regarding the design of the \textit{Global Quantum Game Jam} (Global QGJ). Subsequently, a mixed-methods study was conducted, following a similar approach as in the first part of this study. The questions aimed to reveal possible shortcomings of the event and to map the possible expectations the participants might have related to the event. 

Two consecutive iterations of the Global QGJ were held in 2021 and 2022. The events ran over a single weekend, both lasting from a Friday to the next Sunday with about 48 hours of game development. The event was hosted on a designated Discord server and online streams of opening remarks and talks were publicly available through Twitch.io. Informative materials about quantum physics, game development and quantum games were offered well in advance as an optional resource. In 2022 additional informative talks on these subjects were offered through an online stream on the first day, prior to the event.

The effectiveness of the event's design and format on participants’ experiences were evaluated using questionnaires distributed through the community server that hosted the events. The questionnaire covered topics such as participants' backgrounds in game development and quantum physics, reasons for participating, and reporting of the usage on the provided resources. Quantitative data about participants' age, nationality, gender and backgrounds related to game development and quantum physics were collected on the first day of the event using multiple-choice questions. Qualitative data on open feedback was gathered on the last day of the event through open-ended questions and analysed thematically. The wording and options of these questions were consistent with the questionnaires used in IFQGJ 2020, IndiQ QGJ 2020 and the GN!2021 jam (See Appendix). In 2021 and 2022, participants were also asked to reflect on their learning through the event. Learning about quantum physics was not specifically mentioned in order to see how this topic would be brought up by the participants. 

Observations were also made during the event, focusing on participants' interactions, behaviours, and use of tools and materials. These observations were restricted to the use of the event server, and were not able to follow any communication that may have happened through other media. 
\section{Data}
For the first part of the study, which focused on the events of the IndiQ QGJ 2020, IFQGJ 2020, and the GN!202 expert jam in 2021, a total of 74 responses from 40 individuals were gathered. These responses provided valuable information specifically regarding the participants' backgrounds in game development and their involvement in the study of quantum physics. 

From the 29 responses given on the first day of IndiQ QGJ 2020 and IFQGJ 2020, a majority of participants (18, 62\%) indicated having participated game developing prior to the event, and from them 12 (42.4\%) had participated to a game jam before and four were professionals with no prior game jam experience. Out of the 29, altogether six (20.7\%) claim being professional game developers and six hobbyists. A bit over half, 15 (51.7\%) of the 29 respondents had at least bachelor’s level education on quantum physics, 13 of them being a quantum physics researcher either with a Ph.D. or studying for one. Out of these 15 respondents with a background in university level quantum physics, nine had no previous experience in game development, six had participated to a game jam before, and one was a hobbyist developer.

Although GN!2021 did not specifically target individuals with a background in quantum physics, out of the 9 responses received, one respondent indicated having a high level of formal education or training in the study of quantum physics. Half of the rest of the respondents reported having no prior knowledge of quantum physics and the rest responded having followed news articles or online videos about the subject. On the other hand, out of the 9 respondents only one had no formal training in game development. This respondent was not the one indicating a university level education in quantum physics. Five out of the nine respondents had participated in commercial game productions, this also included the quantum physicist. It can be concluded that the event was able to attract a high level of expertise in game development, but reached only a niche sampling of quantum physics expertise.

17 responses were gathered on the third day of the events of IndiQ QGJ 2020 and IFQGJ 2020. From the open-ended questions, 6 reported having learned something new, and 6 other respondents claimed having "no idea" or feeling a sense of confusion, when specifically asked about their thoughts on quantum physics.

In the context of the IFQGJ, a participant with a high school level understanding of quantum physics expressed that the event was not particularly conducive to learning, stating, \textit{"I ended up using the concepts I remembered the most even if they weren't really that connected to the theme."} In addition, a graphic artist highlighted the challenging nature of a game jam environment and remarked, \textit{"it's very hard for a person that never had to do anything with quantum to learn something in this environment and with this little time."}

Based on the open-ended responses received, it was inferred that incorporating learning on a challenging subject like quantum physics while engaging in game development posed a significant challenge, though the acquisition of new knowledge related to quantum physics was appreciated. While some participants with a background in quantum physics expressed the benefit of attempting to explain quantum concepts to their team members, it was observed during the events that this approach of teaching during game development time might have impeded the overall effectiveness of the game development process. It was also proposed by one respondent that they would prefer joining the opening talks the day before to have had time to carefully listen to them. 

To address this challenge, online materials covering fundamental quantum physics concepts and the general structure of game development were specifically created for the Global QGJ. These resources were made available to participants well in advance, enabling them to familiarise themselves with the materials prior to the event, whenever feasible. In addition for the 2022 edition of the Global QGJ, informative talks on basic concepts related to quantum physics and quantum computing as well as game development were offered to anyone interested prior to the event itself. 

Seven games were developed during the IndiQ QGJ, eight during the IFQGJ and eight in GN! game jam. From the 23 games that were submitted in the events of IndiQ QGJ, IFQGJ and the GN!2021, 13 (56.52\%) credited their creators. 39 individual credentials were marked for these games. Based on observations during the event, the un-credited games may have had at least a number of 12 creators involved in them as several can be suspected to have been solo projects. With respect to this number, it may be seen that the sampling of 40 individuals in the study seems adequate for the purpose of developing the event design further. 

Not all the submitted games provided a separate explanation about the relations to quantum physics or quantum computing on their Itch.io submission page. Out of the 23, five state using a numerical simulation of quantum physical phenomena or quantum computers, and for three games the connection to quantum physics was unclear. Rest of the games had at minimum a thematic reference to quantum physics, one with clear educational aspects.

Based on the data collected and observations made during these events in 2020 and 2021, it became evident that the quantum physics themed game jams were successful in attracting and engaging a participant audience, particularly those with a background in quantum physics rather than game development. Additionally, it was observed that the participants primarily comprised members from the groups affiliated with the event organisers, highlighting the need to refine recruitment strategies to attract a more diverse participant pool. Therefore for the recruitment of the Global QGJ of 2021 and 2022 special care was taken into inviting experienced game developers both for participating in the event as well as for mentors. 
\subsection{Participants of the Global Quantum Game Jam}
From the Global QFJ events of 2021 and 2022 overall a number of 151 responses from 124 individuals were gathered. In both iterations, two questionnaires were distributed. The first questionnaire served also as a registration form for the event, and the second was distributed after game submission on the third day of the event. 

The Global QGJ event in 2021 attracted 42 registrations on the first day, with 11 female, 28 male, 1 choosing the option of “Other” as their answer, and 2 preferring not to disclose their gender. The average age of the study respondents was 28 years with a standard deviation of 7.4 years, with the youngest respondent being 19 years old and the oldest of the age 46 years. The event reached participants from all the continents with the exception of Antarctica. The respondents were mostly Asian (16 respondents out of 42), European (16 respondents out of 42) and African (6 respondents out of 42). Most common nationalities reported were Indian (10 out of 42) and Finnish (9 out of 42).

In 2022, 78 registrations were received, 26 from females, 49 from males, two choosing the option “Other”, and 1 from someone not wanting to disclose their gender. The average age of the respondents was 26 years with a standard deviation of 7.2 years. Youngest of the registered participants were 14 and 15 years old, oldest being 46 years old. The event reached participants from all the continents with the exception of Antarctica. Respondents were mostly Asian (35 respondents out of 78), and European (25 respondents out of 78)). Most common nationalities were Indian (19 respondents out of 78), and Finnish (12 respondents out of 78). One answer out of the 78 did not indicate any specific country.

The Global QGJ events attracted mainly participants with no prior background in game development. Based on the data collected from the two iterations of the Global QGJ, it was found that 53.5\% (54) of the 120 respondents who participated in the initial days of the event reported having no prior experience in game development (See Figure \ref{table:gpexperience00}). There was minimal variation in these numbers between the two iterations (See Appendix). Furthermore, out of the 120 responses received regarding prior participation in game jams, 59,1\% (71) indicated that the Global QGJ was their first experience with game jams. 
\begin{figure}[h]
  \centering
  \caption{The participant background related to experience in game development at the Global QGJ of 2021 and 2022, 120 respondents.}
  \includegraphics[width=0.8\linewidth]{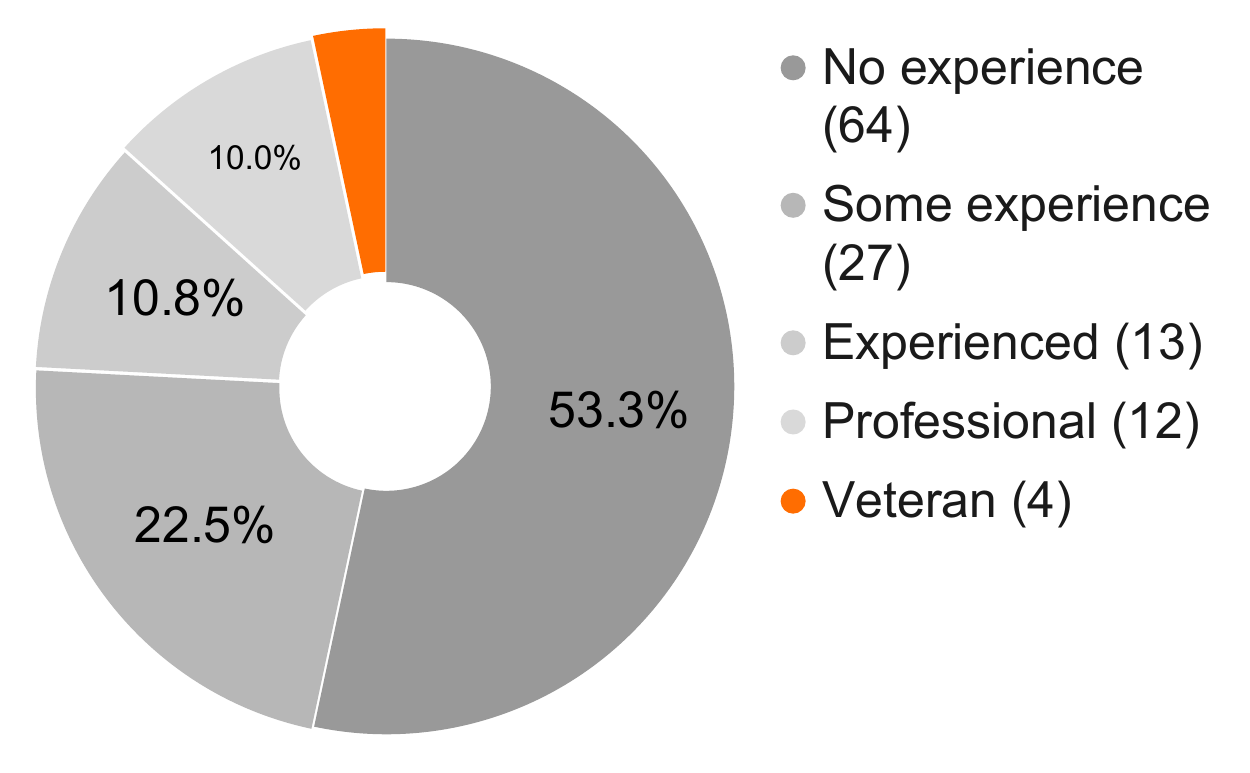}
   \label{table:gpexperience00}
\end{figure}

\begin{figure}[h]
  \centering
  \caption{The background related to  quantum physics of the participants of the Global QGJ of 2021 and 2022, 120 respondents.}
  \includegraphics[width=0.9\linewidth]{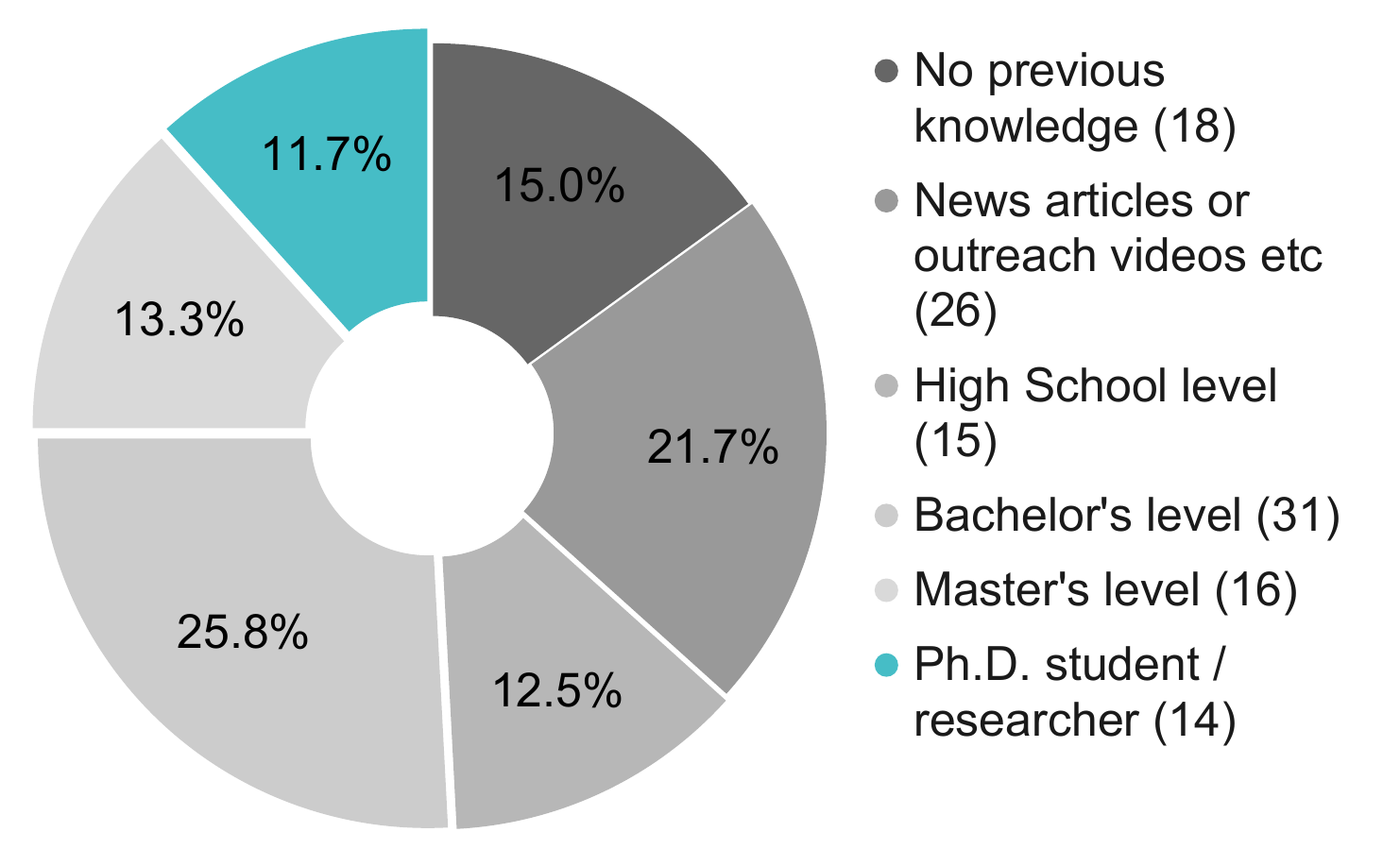}
   \label{table:qpexperience00}
\end{figure}

Regarding the participants' background in relation to quantum physics, a total of 120 responses were received. Slightly more than half of the respondents, specifically 50.83\% (61), reported having taken at least bachelor's level courses in quantum physics. Conversely, only 15.0\% (18) indicated that they had no previous knowledge in this field (See Figure \ref{table:qpexperience00}). 

\subsubsection{Use of provided resources}
Extra resources and information about quantum physics, game development and quantum games were offered through an online folder for the participants of 2021 and 2022 Global QGJs.

120 responses were gathered about the use of the online materials on the first day of the event iterations of 2021 and 2022 of Global QGJ. The majority (78, 65\%) of the respondents were aware of the materials, 18.33\% (22) found them “very useful” (See Appendix). Out of the 31 answers received on the last day of the events, 32.26\% (10) had used the materials during the jam and 22.58\% (7) had used them before the start of the event. This measures to over a half of the respondents finding the materials useful either prior to the event or during it. 

It has been observed from the open-ended answers, that individuals without prior expertise in game development expressed a significant sense of achievement and learning through the process of using new tools and principles in game development. Their learning primarily stemmed from the actual act of developing games. In contrast, no participants specifically mentioned acquiring substantial knowledge in quantum physics, aside from thematic ideas. The burgeoning intrigue surrounding quantum physics and the keenness to deepen one's understanding was reported by many of the respondents as well as other participants through the event platform, thus underlining the importance of low-level introductory materials. For individuals with a formal education in quantum physics, the experience served as a means to address their limitations of understanding and offered an opportunity to fill in these gaps.\\

\subsubsection{The Games}
Seventeen names were credited on the seven games uploaded on the event pages on Itch.io during the event iteration of 2021. The event iteration in 2022 resulted in 18 games with 44 credited individuals. Altogether a number of 25 games have been developed in the studied 2021 and 2022 and all the games included credits of the team members. All the teams behind the submitted games identified the connection to quantum physics through the Itch.io submission form. Based on them only one submission cannot be considered a quantum game through the definition of quantum games \cite{piispanen2023}, and this particular game was not developed by registered or active participants of the QGJ. From the rest, at least a thematic, \textit{perceivable dimension of quantum physics} was reported. In addition, several of the games had a \textit{dimension of scientific purposes} through their intent to serve as an educational game, and through the use of quantum computing simulators, some games also presented a \textit{dimension of quantum technologies} \cite{piispanen2023}. In particular, the use of Qiskit, open-source SDK for quantum computing, was reportedly used in most of such games. \\

For the online QGJ, many respondents reported about challenges related to communication. Still, the aspect of socialising and establishing new connections emerged as a prominent highlight of the event, as indicated by the majority of participants when asked about the most favourable aspects of the event. Overall, from all the responses received on the last day of the event, 30 responses were received to the question of “Would you join a Quantum Game Jam again?”. From them 80\% (24) responded with “Yes”, 16.67\% (5) with “Maybe” and 3.3\% (1) with “No”. 
\section{Guidelines for organising an online Quantum Game Jam} 
Drawing from our experience of organising five Quantum Game Jams (QGJ), three of which were held online, we provide the following set of best practises for organising an online QGJ. We also draw from our experiences of organising over 30 game jams, from which 8 were held online and four as hybrid events.\\

\textbf{Create a code of conduct for a QGJ in advance}. It can be based on an open code provided by organisations such as the International Game Developers Association (IGDA). \textbf{Upon selecting the online platforms for the event, prioritise well-known and user-friendly options, while also keeping the overall number of platforms in minimum}. It should be clearly informed which of them is the main platform to follow. Verify that the selected platform allows for the establishment of multiple team channels. Our experiences are based on using the combination of Discord and Twitch.tv, with Itch.io serving as the platform for registering and publication of games.

To facilitate effective communication and information dissemination, it is essential to \textbf{establish a dedicated channel that provides regular updates about the event}, including schedule changes. This channel serves to guide participants to the appropriate platforms or channels for their engagement, minimising information overload and ensuring they stay informed about the latest developments. Allow only the organisers to comment on the main informative channel and keep the information to minimum. Mentors can play a crucial role in directing participants to the relevant timetable and information channels. 

Furthermore, it is recommended to create specific channels catering to different needs, such as talent-seeking, brainstorming, a general helpdesk, casual conversations, and channels dedicated to participants introducing themselves and their talents and backgrounds based on their respective areas of expertise. This setup benefits both organisers and teams in finding the right individuals to complement their teams.\\

\textbf{Advertise the event well in advance to enable potential participants to plan their schedules accordingly}. Also, reaching out to the right audiences ensures that the event attracts a diverse group of individuals with varied backgrounds and skills, contributing to a well-rounded mix of talents among the participants. By keeping the online platform active well in advance of the event, organisers may help reduce participants' anxiety about using the platform and engaging in social activities. The authors recommend recruiting mentors with expertise on the usage of the platforms, on game development in game jams and on quantum physics.

\textbf{Consider the diverse backgrounds of potential participants}. Specifically, we recommend providing a series of introductory-level talks and online materials that cover the fundamental principles and key features of quantum physics, and the possible simulation tools used, as well as the essential aspects of the game development process, particularly within the context of a time-limited game jam. In particular, example projects using the offered simulation tools are highly recommended to be shared. 

To further support participants, we recommend that the talks be scheduled and recorded prior to the event, and made available for later viewing. While participation in the talks should not be mandatory, they provide an opportunity for participants to better understand the event's theme and improve their chances of creating a game with a connection to quantum physics or quantum computing. \textbf{Provide clear instructions on how to use the online platforms, the event schedule, and general guidelines on how to prepare for the event}.\\

\textbf{Organise registration to the event on the first day of the event} through services like Google Forms. These forms may also help organisers gain information about the backgrounds of the participants, and to check upon participants that might be missing a team, and make them feel more welcomed to socialising and active participation of the platform. For this, we recommend either using real names for both registration and on the main platform, or collecting the nicknames separately. Emphasise also on the registration form, what is the main platform and the main channel to follow during the event.

\textbf{Start the Quantum Game Jam with welcoming words, introductions on topic and provide essential information about the event}. Given that participants are not expected to thoroughly study and internalise any provided materials and lectures prior to the event, it is essential to clearly present critical information during the event's opening. It is equally important to provide a clear presentation of the event structure, the challenge of the time restriction, and emphasise on the collaborative and creative nature of the event. This is done through a streaming platform, which allows for recording of the talks for participants joining from other time zones, and does not limit the number of viewers. The use of platforms like Twitch.tv to stream the talks and the event narrative allows for greater flexibility and broader participation than using platforms such as Discord for streaming. Test the sound quality of the stream well beforehand prior to the event.

After the presentations, the event's theme is revealed, directing the participants towards developing new game ideas from scratch, rather than relying on pre-planned concepts. \\

\textbf{Facilitate the brainstorming session by encouraging participants to contribute diverse suggestions}. Activate the use of the text-based platform with the event stream by hosting a guided, collective brainstorming session, and be informative also about the time reserved for the activity. A dedicated brainstorming channel, moderated and monitored via the streaming platform, facilitates this process. By combining supervision through live streaming and text-based communication, the organiser can stimulate idea generation and assist participants in expanding upon their ideas. This is also important when overseeing and guiding the formation of balanced teams. 

Following the brainstorming session, guide participants to pitch their game concepts on the brainstorming channel. Provide a structured format via the live stream and directly on the channel. Encourage the use of drawings, mood boards, videos or other possible references. Guide towards editing the original pitch instead posting a similar one, and encourage the use of possible platform provided reactions on these pitches. Present and comment on the pitches through the stream. 

Organisers should start the creation of team channels for the proposed game ideas. The pitch may be copied to start the discussion on the channel and the further building of the idea collectively. Encourage participants to join these channels by introducing themselves and the possible talents they may provide for realising the game. Offer also the possibility for voice channels. This approach streamlines the process and encourages participants to engage in further collaborative discussions. Google Sheets or similar tools may be used to collect information on team formations, including the team name, a contact person, their contact information, the roles each participant can fill, and information on missing expertise. \\

\textbf{Guide participants to agree on sequent meetings through the event and to use the main platform of the event for communication}. In order to facilitate mentoring and other forms of assistance for the participants, it is crucial that team communication takes place within the primary platform of the event. Additionally, it is advantageous for the team to engage in early discussions regarding the tools and documentation utilised, promoting a seamless collaboration.

\textbf{Provide mentors with expertise in quantum physics and game development for the participants throughout the event}. This resource is especially helpful to teams lacking specific skills or members with certain backgrounds. Mentors should coordinate a schedule throughout the jam to ensure that there is at least one quantum physics expert and one game development expert available for the teams at any given time, or whenever possible.  By setting server specific roles for mentors allows them to be tagged in comments related to specific problems and questions the participants might have. Such server-specific roles may be for example quantum physics expert, game design expert, programming expert etc.

The majority of the jam's duration is spent within teams during the development phase. \textbf{Organise social activities on the online platforms during the development phase}, such as additional talks and discussion sessions, mid-milestone reporting on project status via Zoom calls or the main platform, and a dedicated casual chat channel. This can facilitate team cohesion and communication. Dedicated casual channels and social activities can also be useful for recruiting project testers for the teams.

\section{Discussion}
The primary purpose of the surveys was to enhance the event through feedback and suggestions. The feedback received has been instrumental in the development of the QGJ events on a global level. It has been used to emphasise the positive aspects and address the identified challenges. Significant updates to the event design include prioritising the creative aspects of game development. This was achieved by supplementing the event with online resources on game development, quantum physics, and event structure, as well as incorporating these elements into the opening presentations. Moreover, informative sessions were introduced before the event, and casual talks on the subject matter were held during the second day. \\

Quantum Game Jams (QGJ) have been able to produce prototypes of quantum games that have since been developed into citizen science games, educational games and art installations \cite{kultima2021qgj, piispanen2023history}. It needs to be noted that the teams behind these successful prototypes have usually had a great deal of expertise on quantum physics, game development and are also familiar with the concept of game jams and the restrictions these events impose. Motivating people to join in sequential QGJs may provide an advantage to the developers in producing novel games with meaningful connections to quantum physics.

Although the Global QGJ has received numerous expressions of gratitude and praise through the feedback forms used in this study, it is worth noting that few of the respondents without previous background in game jams found the task of creating a quantum physics-related game challenging. The most challenging part of these online events has been in facilitating team building and making sure everyone interested finds a team or a meaningful solo project they are able to commit to for the duration of the jam. We find it important to underline that the aim of a QGJ is not to present a highly polished and ready game, but to come together with other people to work on a creative task. We also hypothesise that respondents may have been hesitant to provide constructive criticisms due to the risk of being identified through their answers. 

A respondent with no previous background in quantum physics or game development prior to the event summarised the main take-away we’d also say is the value of QGJs for game developers: “[Quantum physics is] \textit{something too difficult and abstract to be understood in 3 days, but something strange enough to become a game idea.}” Quantum physics gives the opportunity to challenge oneself and stretch the limits of imagination. \\

Unlike traditional game jams, which often prioritise minimal lectures and presentations \cite{kultima2016fgj}, a science game jam necessitates a certain level of subject matter knowledge to establish a solid foundation for the development of impactful science games. In addition, it is important to emphasise the equality of team members and their shared commitment to a common goal, rather than serving only the interests of any individual member. The authors have recognised the significance of establishing a strong foundation in understanding game development as a creative process in science game jams, and this principle extends to promoting teamwork and collaboration among all team members.

It is evident that individuals without a prior background in quantum physics often face challenges in determining their initial steps towards a quantum physics-related project. Quantum physics is regarded as an intriguing yet intricate subject, and there is often an expectation that individuals should possess a solid understanding of complex concepts before delving into the problem itself. Through years of organising QGJs, quantum game courses, personal projects, and interactive quantum art development, the authors have recognised the significance of providing an example project accompanied by a numerical simulation of a quantum mechanical object or a quantum computing simulator such as Qiskit. Integrating these tools and ensembles can significantly lower the entry barrier for individuals to embark on their creative endeavours.

Moreover, comprehensive documentation and instructions regarding the functionality of the simulator, technical requirements, and potential installation processes are crucial. In the context of events like the QGJ or quantum game development courses, the availability of on-site or online technical support is highly desirable. This support should encompass assistance with simulation tools and other technical necessities to ensure a smooth and efficient experience for participants.\\

The most prominent distinction between the on-site QGJs and online quantum game jams is the social aspect. On-site game jams in general have offered fruitful ground for socialising, networking and even for developing friendships both within the members of a single team but most distinctively between the teams \cite{reng2013}. At on-site QGJs we have witnessed a multitude of ways people have interacted with each other in the same room; technical help has been offered between the teams, testing of games and most importantly discussions not related to the design problems at hand, but instead providing a valuable break. It is important to facilitate such opportunities also through the online events, as the social aspect has arisen as a prominent positive factor related to the events studied in this paper. 

Another notable distinction is that local event venues often impose restrictions on participant numbers, making it necessary to control attendance through a limited number of available tickets. However, by carefully planning the online platforms utilised, this limitation can be overcome. Given that the Global QGJ has thus far maintained a moderate scale, the implementation of a separate registration system has not been considered indispensable. An online game jam already presents a significant learning curve in terms of platform usage and notifications, rendering an additional registration system unnecessary during the event. Consequently, the decision was made to eliminate the use of a pre-registering platform in 2022 to streamline the participant experience.

It is important to guide the participants to communicate through the common online platform, especially when they are not familiar with each other from before the event. Our event follows the IGDA Finland Code of Conduct and has also server-specific rules underlining the zero tolerance policy towards any type of harassment/abuse/toxic behaviour. If a participant is to behave in a way that offends any other participant, the organiser is able to address only the behaviour they are able to witness. To ensure equitable distribution of credit and effective conflict resolution, it is strongly recommended to utilise a designated server for all team communication.\\

Comprehensively understanding the motivations and resources individuals bring to an event, such as this, involves exploring both participation and withdrawal. Focusing solely on those who completed the event risks overlooking valuable insights into the reasons why some participants withdrew, despite expressing motivation to continue. Therefore, it would be important to engage with these individuals to gain valuable feedback that can help improve future iterations of the event. \\

Through joining a QGJ and facing the challenges of communicating one’s expertise for their team offers the value of seeing a familiar subject of matter through new lenses. At best, it brings clarity to the things that might be seen as obstacles and brings to the attention possible misconceptions people might have and also offers a valuable prototype of a game that may be used for different outreach and educational purposes. The authors aim to continue expanding the QGJ event each year and are looking forward to any enthusiasts reading this paper in joining the event, either as a participant or even as a volunteer in helping organise these events. 
\section{Conclusions}
In this paper we have investigated altogether five iterations of online, quantum physics themed game jams, \textit{Quantum Game Jams} (QGJ). As the results of this paper, the feedback and lessons learned were used in the development of the series of the Global QGJ and lead to guidelines for organising online science game jams through the example of online QGJs. 

In an online QGJ, where many participants may have limited experience in game development, it is crucial to actively foster the communal nature of brainstorming and game development, ensuring that all participants have the opportunity to actively engage in the creative process. This inclusive approach encourages active participation from all attendees and promotes a collaborative and supportive environment. Equally important is the provision of well-crafted educational materials on quantum physics and clear instructions regarding the available development tools for quantum games, as well as guidance on navigating the event platform(s).

\begin{acks}
Piispanen would like to acknowledge their research has been funded by the use of Academy of Finland PROFI funding under the Academy decision number 318937 and by the Vilho, Yrjö and Kalle Väisälä Foundation of the Finnish Academy of Science and Letters.
\end{acks}

\bibliographystyle{ACM-Reference-Format}
\bibliography{references}


\begin{thebibliography}{23}


\ifx \showCODEN    \undefined \def \showCODEN     #1{\unskip}     \fi
\ifx \showDOI      \undefined \def \showDOI       #1{#1}\fi
\ifx \showISBNx    \undefined \def \showISBNx     #1{\unskip}     \fi
\ifx \showISBNxiii \undefined \def \showISBNxiii  #1{\unskip}     \fi
\ifx \showISSN     \undefined \def \showISSN      #1{\unskip}     \fi
\ifx \showLCCN     \undefined \def \showLCCN      #1{\unskip}     \fi
\ifx \shownote     \undefined \def \shownote      #1{#1}          \fi
\ifx \showarticletitle \undefined \def \showarticletitle #1{#1}   \fi
\ifx \showURL      \undefined \def \showURL       {\relax}        \fi
\providecommand\bibfield[2]{#2}
\providecommand\bibinfo[2]{#2}
\providecommand\natexlab[1]{#1}
\providecommand\showeprint[2][]{arXiv:#2}

\bibitem[Archer(2022)]%
        {archer2022}
\bibfield{author}{\bibinfo{person}{Noora Archer}.}
  \bibinfo{year}{2022}\natexlab{}.
\newblock \emph{\bibinfo{title}{{Visual design of quantum physics – Lessons
  learned from nine gamified and artistic quantum physics projects}}}.
\newblock Master's thesis. \bibinfo{school}{Aalto University. School of Arts,
  Design and Architecture}.
\newblock
\urldef\tempurl%
\url{http://urn.fi/URN:NBN:fi:aalto-202301291591}
\showURL{%
\tempurl}


\bibitem[Arya et~al\mbox{.}(2013)]%
        {arya2013}
\bibfield{author}{\bibinfo{person}{Ali Arya}, \bibinfo{person}{Jeff Chastine},
  \bibinfo{person}{Jon Preston}, {and} \bibinfo{person}{Allan Fowler}.}
  \bibinfo{year}{2013}\natexlab{}.
\newblock \showarticletitle{An international study on learning and process
  choices in the global game jam}.
\newblock \bibinfo{journal}{\emph{International Journal of Game-Based Learning
  (IJGBL)}} \bibinfo{volume}{3}, \bibinfo{number}{4} (\bibinfo{year}{2013}),
  \bibinfo{pages}{27--46}.
\newblock


\bibitem[Fowler et~al\mbox{.}(2013)]%
        {fowler2013learn}
\bibfield{author}{\bibinfo{person}{Allan Fowler}, \bibinfo{person}{Foaad
  Khosmood}, \bibinfo{person}{Ali Arya}, {and} \bibinfo{person}{Gorm Lai}.}
  \bibinfo{year}{2013}\natexlab{}.
\newblock \showarticletitle{The global game jam for teaching and learning.}
\newblock \bibinfo{journal}{\emph{Proccedings of the 4th Annual Conference on
  Computing and Information Technology Research and Education New Zealand}}
  (\bibinfo{year}{2013}), \bibinfo{pages}{28--34}.
\newblock


\bibitem[Ho et~al\mbox{.}(2014)]%
        {ho14}
\bibfield{author}{\bibinfo{person}{Xavier Ho}, \bibinfo{person}{Martin
  Tomitsch}, {and} \bibinfo{person}{Tomasz Bednarz}.}
  \bibinfo{year}{2014}\natexlab{}.
\newblock \showarticletitle{Game Design Inspiration in Global Game Jam}. In
  \bibinfo{booktitle}{\emph{DiGRA 2014: What is Game Studies in Australia?}}
  \bibinfo{publisher}{Australia: DiGRA}.
\newblock


\bibitem[Koskinen(2021)]%
        {koskinen2021}
\bibfield{author}{\bibinfo{person}{Elina Koskinen}.}
  \bibinfo{year}{2021}\natexlab{}.
\newblock \showarticletitle{Pizza and Coffee Make a Game Jam - Learnings From
  Organizing an Online Game Development Event}. In
  \bibinfo{booktitle}{\emph{Sixth Annual International Conference on Game Jams,
  Hackathons, and Game Creation Events}} (Montreal, Canada)
  \emph{(\bibinfo{series}{ICGJ 2021})}. \bibinfo{publisher}{Association for
  Computing Machinery}, \bibinfo{address}{New York, NY, USA},
  \bibinfo{pages}{74–77}.
\newblock
\showISBNx{9781450384179}
\urldef\tempurl%
\url{https://doi.org/10.1145/3472688.3472699}
\showDOI{\tempurl}


\bibitem[Kultima(2015)]%
        {kultima2015}
\bibfield{author}{\bibinfo{person}{Annakaisa Kultima}.}
  \bibinfo{year}{2015}\natexlab{}.
\newblock \showarticletitle{Defining Game Jam}. In
  \bibinfo{booktitle}{\emph{Proceedings of the 10th International Conference on
  the Foundations of Digital Games}}.
\newblock


\bibitem[Kultima(2018)]%
        {kultima2018}
\bibfield{author}{\bibinfo{person}{Annakaisa Kultima}.}
  \bibinfo{year}{2018}\natexlab{}.
\newblock \showarticletitle{Design Values of Game Jam Organizers - Case: Global
  Game Jam 2018 in Finland}. In \bibinfo{booktitle}{\emph{Proceedings of
  International Conference on Game Jams, Hackathons and Game Creation Events}}.
  \bibinfo{publisher}{ACM}, \bibinfo{address}{San Francisco, CA, USA}.
\newblock


\bibitem[Kultima et~al\mbox{.}(2016)]%
        {kultima2016fgj}
\bibfield{author}{\bibinfo{person}{Annakaisa Kultima}, \bibinfo{person}{Alha
  Kati}, {and} \bibinfo{person}{Nummenmaa Timo}.}
  \bibinfo{year}{2016}\natexlab{}.
\newblock \showarticletitle{Building Finnish Game Jam Community through
  Positive Social Facilitation}. In \bibinfo{booktitle}{\emph{Academic
  Mindtrek}}. \bibinfo{publisher}{ACM}.
\newblock


\bibitem[Kultima et~al\mbox{.}(2021a)]%
        {kultima2021expert}
\bibfield{author}{\bibinfo{person}{Annakaisa Kultima}, \bibinfo{person}{Solip
  Park}, \bibinfo{person}{Ville Kankainen}, \bibinfo{person}{Riikka Aurava},
  \bibinfo{person}{Laura Piispanen}, {and} \bibinfo{person}{Tomi Kauppinen}.}
  \bibinfo{year}{2021}\natexlab{a}.
\newblock \showarticletitle{Expert-Driven (Online) Game Jams for (Game) Design
  Education} \emph{(\bibinfo{series}{ICGJ 2021})}.
  \bibinfo{publisher}{Association for Computing Machinery},
  \bibinfo{address}{New York, NY, USA}, \bibinfo{pages}{64–68}.
\newblock
\showISBNx{9781450384179}
\urldef\tempurl%
\url{https://doi.org/10.1145/3472688.3472697}
\showDOI{\tempurl}


\bibitem[Kultima et~al\mbox{.}(2021b)]%
        {kultima2021qgj}
\bibfield{author}{\bibinfo{person}{Annakaisa Kultima}, \bibinfo{person}{Laura
  Piispanen}, {and} \bibinfo{person}{Miikka Junnila}.}
  \bibinfo{year}{2021}\natexlab{b}.
\newblock \showarticletitle{Quantum Game Jam – Making Games with Quantum
  Physicists}.
\newblock  (\bibinfo{year}{2021}), \bibinfo{pages}{134–144}.
\newblock
\showISBNx{9781450385145}
\urldef\tempurl%
\url{https://doi.org/10.1145/3464327.3464349}
\showURL{%
\tempurl}


\bibitem[Lai et~al\mbox{.}(2021)]%
        {lai2021}
\bibfield{author}{\bibinfo{person}{Gorm Lai}, \bibinfo{person}{Annakaisa
  Kultima}, \bibinfo{person}{Foaad Khosmood}, \bibinfo{person}{Johanna Pirker},
  \bibinfo{person}{Allan Fowler}, \bibinfo{person}{Ilaria Vecchi},
  \bibinfo{person}{William Latham}, {and} \bibinfo{person}{Frederic
  Fol~Leymarie}.} \bibinfo{year}{2021}\natexlab{}.
\newblock \showarticletitle{Two decades of game jams}. In
  \bibinfo{booktitle}{\emph{Sixth Annual International Conference on Game Jams,
  Hackathons, and Game Creation Events}}. \bibinfo{pages}{1--11}.
\newblock


\bibitem[Laiti et~al\mbox{.}(2020)]%
        {laiti2020}
\bibfield{author}{\bibinfo{person}{Outi Laiti}, \bibinfo{person}{Sabine
  Harrer}, \bibinfo{person}{Satu Uusiautti}, {and} \bibinfo{person}{Annakaisa
  Kultima}.} \bibinfo{year}{2020}\natexlab{}.
\newblock \showarticletitle{Sustaining intangible heritage through video game
  storytelling - the case of the Sami Game Jam}.
\newblock \bibinfo{journal}{\emph{International Journal of Heritage Studies}}
  \bibinfo{volume}{0}, \bibinfo{number}{0} (\bibinfo{year}{2020}),
  \bibinfo{pages}{1--16}.
\newblock
\urldef\tempurl%
\url{https://doi.org/10.1080/13527258.2020.1747103}
\showDOI{\tempurl}
\showeprint{https://doi.org/10.1080/13527258.2020.1747103}


\bibitem[Musil et~al\mbox{.}(2010)]%
        {musil2010}
\bibfield{author}{\bibinfo{person}{Juergen Musil}, \bibinfo{person}{Angelika
  Schweda}, \bibinfo{person}{Dietmar Winkler}, {and} \bibinfo{person}{Stefan
  Biffl}.} \bibinfo{year}{2010}\natexlab{}.
\newblock \showarticletitle{Synthesized essence: what game jams teach about
  prototyping of new software products}. In
  \bibinfo{booktitle}{\emph{Proceedings of the 32nd ACM/IEEE International
  Conference on Software Engineering-Volume 2}}. \bibinfo{pages}{183--186}.
\newblock


\bibitem[Piispanen(2017)]%
        {quantumgames}
\bibfield{author}{\bibinfo{person}{Laura Piispanen}.}
  \bibinfo{year}{2017}\natexlab{}.
\newblock \bibinfo{title}{List of Quantum Games}.
\newblock \bibinfo{howpublished}{\url{https://kiedos.art/quantum-games-list/}}.
\newblock
\newblock
\shownote{Accessed: 2023-06-11}.


\bibitem[Piispanen(2023)]%
        {piispanen2023projects}
\bibfield{author}{\bibinfo{person}{Laura Piispanen}.}
  \bibinfo{year}{2023}\natexlab{}.
\newblock \showarticletitle{Designing Quantum Games and Quantum Art for
  Exploring Quantum Physics}.
\newblock \bibinfo{journal}{\emph{IEEE Conference on Games (CoG)}}
  (\bibinfo{date}{08} \bibinfo{year}{2023}).
\newblock


\bibitem[Piispanen et~al\mbox{.}(2023a)]%
        {piispanen2023history}
\bibfield{author}{\bibinfo{person}{Laura Piispanen}, \bibinfo{person}{Edward
  Morrell}, \bibinfo{person}{Marcel Pfaffhauser}, \bibinfo{person}{Solip Park},
  {and} \bibinfo{person}{Annakaisa Kultima}.} \bibinfo{year}{2023}\natexlab{a}.
\newblock \showarticletitle{History of Quantum Games}.
\newblock \bibinfo{journal}{\emph{IEEE Conference on Games (CoG)}}
  (\bibinfo{date}{08} \bibinfo{year}{2023}).
\newblock


\bibitem[Piispanen et~al\mbox{.}(2023b)]%
        {piispanen2023}
\bibfield{author}{\bibinfo{person}{Laura Piispanen}, \bibinfo{person}{Marcel
  Pfaffhauser}, \bibinfo{person}{Kultima Annakaisa}, {and}
  \bibinfo{person}{James Wootton}.} \bibinfo{year}{2023}\natexlab{b}.
\newblock \showarticletitle{Defining Quantum Games}.
\newblock  (\bibinfo{date}{05} \bibinfo{year}{2023}).
\newblock
\urldef\tempurl%
\url{https://doi.org/10.48550/arXiv.2206.00089}
\showDOI{\tempurl}


\bibitem[Preston et~al\mbox{.}(2012)]%
        {preston2012}
\bibfield{author}{\bibinfo{person}{Jon Preston}, \bibinfo{person}{Jeff
  Chastine}, \bibinfo{person}{Casey O'Donnell}, \bibinfo{person}{Tony Tseng},
  {and} \bibinfo{person}{Blair MacIntyre}.} \bibinfo{year}{2012}\natexlab{}.
\newblock \showarticletitle{Game Jams: Community, Motivations, and Learning
  among Jammers}.
\newblock \bibinfo{journal}{\emph{International Journal of Game-Based
  Learning}}  \bibinfo{volume}{2} (\bibinfo{date}{07} \bibinfo{year}{2012}),
  \bibinfo{pages}{51--70}.
\newblock
\urldef\tempurl%
\url{https://doi.org/10.4018/ijgbl.2012070104}
\showDOI{\tempurl}


\bibitem[Reng et~al\mbox{.}(2013)]%
        {reng2013}
\bibfield{author}{\bibinfo{person}{Lars Reng}, \bibinfo{person}{Henrik
  Schoenau-Fog}, {and} \bibinfo{person}{Lise~Busk Kofoed}.}
  \bibinfo{year}{2013}\natexlab{}.
\newblock \showarticletitle{The motivational power of game communities-engaged
  through game jamming}. In \bibinfo{booktitle}{\emph{Proceedings of the 8th
  International Conference on the Foundations of Digital Games}}.
  \bibinfo{pages}{14--17}.
\newblock


\bibitem[Scott and Ghinea(2013)]%
        {scott2013}
\bibfield{author}{\bibinfo{person}{Michael Scott} {and}
  \bibinfo{person}{Gheorghita Ghinea}.} \bibinfo{year}{2013}\natexlab{}.
\newblock \showarticletitle{Promoting Game Accessibility: Experiencing and
  Induction on Inclusive Design Practice at the Global Games Jam}.
\newblock \bibinfo{journal}{\emph{The Inaugural Workshop on the Global Game Jam
  (GGJ’13) The 8th International Conference on the Foundations of Digital
  Games 14}}.
\newblock


\bibitem[Skult and Smed(2022)]%
        {skult2022}
\bibfield{author}{\bibinfo{person}{Natasha Skult} {and} \bibinfo{person}{Jouni
  Smed}.} \bibinfo{year}{2022}\natexlab{}.
\newblock \bibinfo{booktitle}{\emph{The Marriage of Quantum Computing and
  Interactive Storytelling}}.
\newblock \bibinfo{pages}{191--206}.
\newblock
\showISBNx{978-3-030-81537-0}
\urldef\tempurl%
\url{https://doi.org/10.1007/978-3-030-81538-7_13}
\showDOI{\tempurl}


\bibitem[Turner et~al\mbox{.}(2013)]%
        {turner2013}
\bibfield{author}{\bibinfo{person}{Truna Aka~J. Turner}, \bibinfo{person}{Lubi
  Thomas}, {and} \bibinfo{person}{Cameron Owen}.}
  \bibinfo{year}{2013}\natexlab{}.
\newblock \showarticletitle{Living the indie life: mapping creative teams in a
  48 hour game jam and playing with data}. In \bibinfo{booktitle}{\emph{IE
  '13}}.
\newblock


\bibitem[Zook and Riedl(2013)]%
        {zook2013}
\bibfield{author}{\bibinfo{person}{Alexander Zook} {and}
  \bibinfo{person}{Mark~O Riedl}.} \bibinfo{year}{2013}\natexlab{}.
\newblock \showarticletitle{Game conceptualization and development processes in
  the global game jam}. In \bibinfo{booktitle}{\emph{Workshop proceedings of
  the 8th international conference on the foundations of digital games}},
  Vol.~\bibinfo{volume}{5}.
\newblock


\end{thebibliography}

\appendix
\begin{table}[h]
  \centering
  \caption{The background related to formal training in quantum physics of the participants of the Global QGJ of 2021 and 2022, number of respondents. The experience the participants may have had related to quantum physics was categorised in a linear matter in reference to formal training.}
  \includegraphics[width=0.9\linewidth]{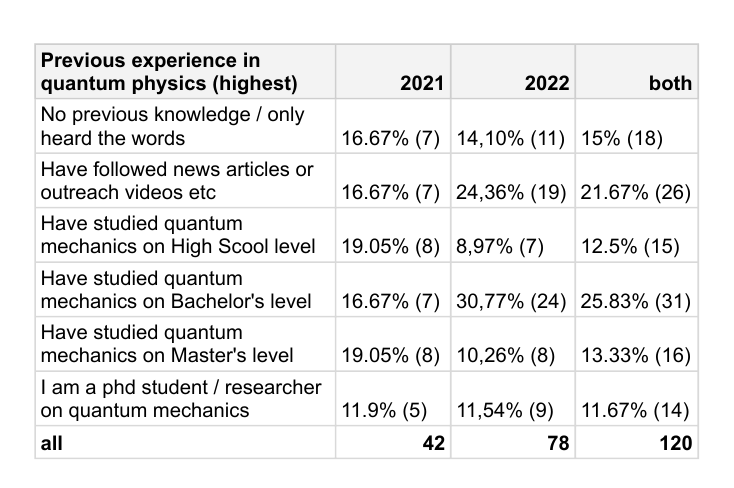}
   \label{table:qpexperience}
\end{table}
\begin{table}[h]
  \centering
   \caption{The formal training related to game making of the participants of the Global QGJ of 2021 and 2022, number of respondents.}
  \includegraphics[width=0.9\linewidth]{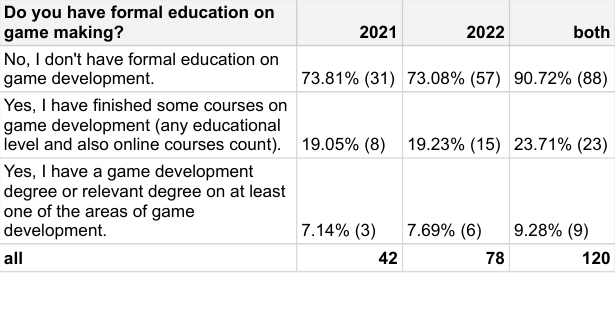}
  \label{table:gamemakingformal}
\end{table}

\begin{table}[h]
  \centering
   \caption{The background related to game making of the participants of the Global QGJ of 2021 and 2022, number of respondents. The information the participants have offered about their formal training in game making, number of commercial games they have been credited for, and with relation to prior game jam participation has been combined and then categorised. Number of answers from 120 responses.}
  \includegraphics[width=0.9\linewidth]{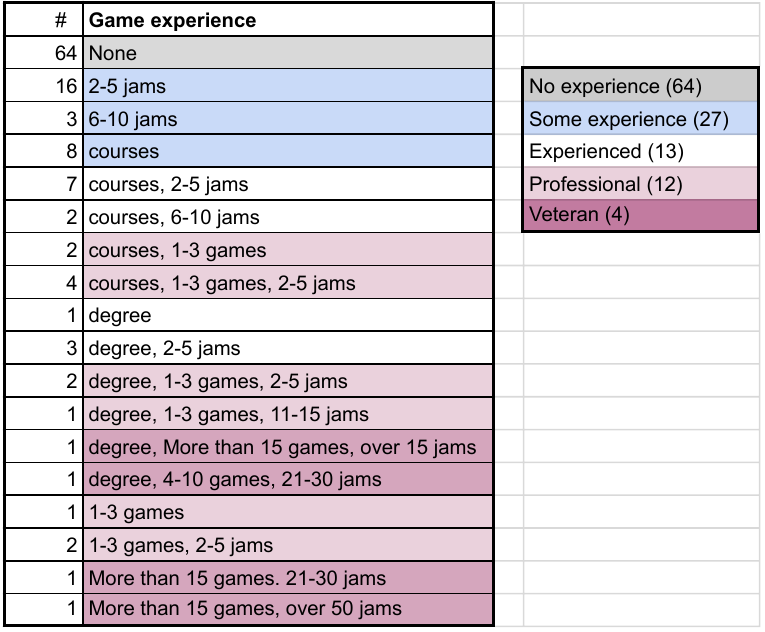}
  \label{table:gamemakingback}
\end{table}

\begin{table}[h]
  \centering
  \caption{The use of the provided online materials during the Global QGJ of 2021 and 2022.}
  \includegraphics[width=0.9\linewidth]{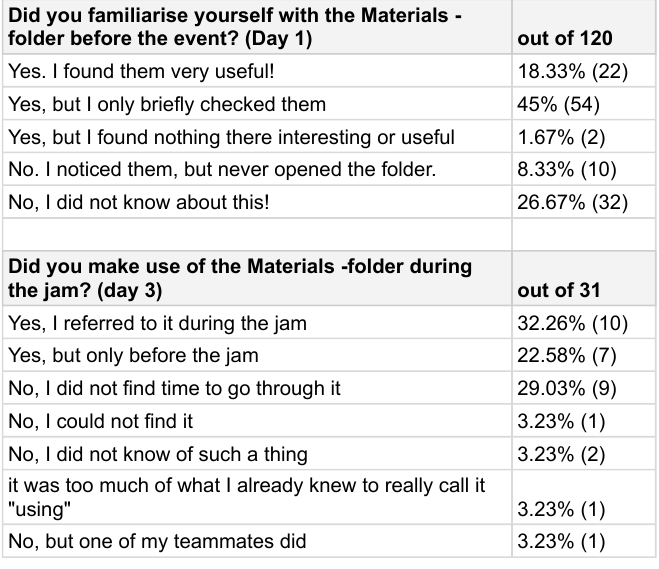}
  \label{table:materials}
\end{table}

\begin{table*}[h]
\centering
\label{table:questions}
\caption{List of questions asked during the Quantum Game Jams of 2020 and Global Game Jam 2021 and 2022. Here MC refers to "multiple-choice", SC to "single-choice" and OE to "open-ended". The "QGJs 2020" refers to the IndiQ Quantum Game Jam in October 2020 and the Pisa Internet Festival Quantum Game Jam in October 2020. GN!2021 refers to the fourth of the Games Now! expert jams in 2021, that had a quantum physics theme. GQGJ stands for Global Quantum Game Jam. For GN!2021 a single set of questions was distributed to the participants after the game jam. For the other events, a questionnaire was distributed on the first day of the event (day 1), and another on the third day of the event, after game submission (day3). }
\begin{tabular}{|l|l|l|}
\hline
\textbf{Questions on day 1:} & \textbf{Type} & \textbf{Event} \\
\hline
\hline
Previous experience in quantum physics & MC & QGJs 2020, GN!2021, GQGJ2021, GQGJ2022 \\
\hline
Previous experience in game development & MC & QGJs 2020\\
\hline
Do you have formal education on game making? & SC & GN!2021, GQGJ2021, GQGJ2022 \\
\hline
How many game jams have you participated (including this jam)?& SC & GN!2021, GQGJ2021, GQGJ2022 \\
\hline
Did you familiarise yourself with the Materials-folder before the event? & SC & GQGJ2021, GQGJ2022 \\
\hline
Write down any thoughts and feelings related to quantum physics. & OE & QGJs 2020, GQGJ2021, GQGJ2022 \\
\hline
Your age & MC & GQGJ2021, GQGJ2022 \\
\hline
Your gender & MC & GQGJ2021, GQGJ2022 \\
\hline
Your nationality & OE & GQGJ2021, GQGJ2022 \\
\hline
\hline
\textbf{Questions on day 3:} & \textbf{Type} & \textbf{Event} \\
\hline
\hline
Any thoughts on quantum mechanics now?  & OE & QGJs 2020, GQGJ2021, GQGJ2022 \\
\hline
Reflect upon on what you learned during this jam. & OE & GN!2021, GQGJ2021, GQGJ2022 \\
\hline
What were the most important features of the jam for your learning? & OE & GN!2021, GQGJ2021, GQGJ2022 \\
\hline
How did you feel about the quantum game jam experience? & OE & QGJs 2020 \\
\hline
Any comments related to the event itself? & OE & QGJs 2020 \\
\hline
What was the best part of the jam for you?& OE & GQGJ2021, GQGJ2022 \\
\hline
What would you have hoped to have happened or organised differently? & OE &GQGJ2021, GQGJ2022 \\
\hline
Did you make use of the Materials-folder during the jam? & MC & GQGJ2021, GQGJ2022 \\
\hline
\end{tabular}
\end{table*}

\end{document}